\documentclass[preprint,pra,showpacs]{revtex4}
\usepackage{amscd,fancyhdr}
\usepackage{amsfonts}
\usepackage[mathscr]{euscript}
\usepackage{pstricks}
\usepackage{pst-node}
\usepackage{mathrsfs}                                                                
\usepackage{color}                                                                   
\usepackage{float,epsfig}
\usepackage{graphicx}                                                                
\usepackage{dcolumn}                                                                 
\usepackage{bm}                                                                      
\usepackage{amsmath,amssymb,amsthm,mathrsfs}
\usepackage[colorlinks,citecolor=green,linkcolor=blue,anchorcolor=blue]{hyperref}    
\usepackage[perpage]{footmisc}
\begin{document}
\title{Hawking radiation as tunneling and the unified
first law of thermodynamics at the apparent horizon in the FRW
universe\footnote{Supported by National Natural Science Foundation
of China under Grant No.~10875060 and Natural Science Foundation of
Shaanxi Education Bureau of China under Grant No.~07JK394. And Feng
J. is supported by NWU Graduate Cross-discipline Funds (08YJC24).}}

\author{Ke-Xia Jiang$^{a,}$\footnote{E-mail address:
kexiaJiang@126.com, kexiachiang@gmail.com}, San-Min Ke$^b$, Dan-Tao
Peng$^a$ and Jun Feng$^a$}

\affiliation{$^a$Institute of Modern Physics, Northwest University,
Xi’an 710069, China\\
$^b$College of Science, Chang'an  University, Xi'an 710064, China}

\renewcommand{\thesection}{\arabic{section}}
\renewcommand{\thesubsection}{\thesection\arabic{subsection}}
\makeatletter
\def\@hangfrom@section#1#2#3{\@hangfrom{#1#2#3}}
\makeatother
\begin{abstract}
Relations between the tunneling rate and the unified first law of
thermodynamics at the apparent horizon of the FRW universe are
investigated. The tunneling rate arises as a consequence of the
unified first law of thermodynamics in such a dynamical system. The
analysis shows obviously how the tunneling is intimately connected
with the unified first law of thermodynamics through the principle
of conservation of energy.
\end{abstract}
\date{\today}
\maketitle
Since Hawking's original work~\cite{Hawking}, Hawking radiation has
been widely discussed for various interests. Several derivations of
Hawking radiation present in literatures~\cite{HartleHGCF}.
Recently, a semi-classical tunneling method~\cite{WK,Parikh,
Angheben, SSP}, attracts many people's attention. The main
ingredient of this method is the consideration of energy conversion
in tunneling of a thin sell from the hole. Many
calculations~\cite{Many cal.} have been investigated by using this
method, and the approach works perfectly. More recently, the general
analysis~\cite{HZZ, Sarkar, Pilling, zhangcai} found this method
gives an interesting result: the tunneling rate $\Gamma \sim
e^{\triangle S}$ arises as a consequence of the first law of
thermodynamics for horizons holds the form, $dE=TdS-PdV$.

However, most analysis of Hawking radiation effect based on static
background spacetimes. There exist an event horizon, which is a
global concept can be used to define the Hawking temperature.
Locally, it is not clear whether there is an event horizon
associated with a certain dynamical spacetime and this causes the
difficulty to discuss in a dynamical situation. Recently, Hayward
\emph{et al.}~\cite{Haywardcvnz} proposed a locally defined Hawking
temperature for dynamical black holes where the
Parikh-Wilczek~\cite{Parikh} tunneling method is used.

Interestingly~\cite{Caietc}, on the apparent horizon of the
Friedmann-Robertson-Walker(FRW) universe, which is a dynamical
system, the Friedmann equation can be rewritten as a thermodynamical
identity, $dE_{h}=TdS+ W dV$. Further, using the tunneling method,
Cai \emph{et al.}~\cite{caicaohu} recently proved that there does
exist Hawking radiation associated with the locally defined apparent
horizon of the FRW universe. The Hawking temperature is measured by
an observer with the Kodama vector~\cite{kodamamsr} inside the
horizon.

To establish thermodynamics of dynamical spacetimes and to show how
it is related with gravity are important problems in General
Relativity. Understanding Hawking radiation is one of the key issues
in steps toward this aim. In this paper, we would like to extend the
work in Ref.~\cite{Sarkar} to the FRW universe, and investigate how
the tunneling is intimately connected with the (unified) first law
of thermodynamics through the principle of conservation of energy in
such a dynamical system, and whether the result that the tunneling
formulas arises as a consequence of the first law of thermodynamics
for horizons holds the form, $dE_{h}=TdS-PdV$, is still reasonable
or not. Indeed, our analysis finally shows that, from the unified
first law of thermodynamics at the apparent horizon of the FRW
universe, one can also get the tunneling formula $\Gamma \sim
e^{\triangle S}$.

Throughout the paper, we take the unit convention $G = c = k_{B}
=\hbar= 1$.

Let's start from the unified first law of thermodynamics at the
apparent horizons in the FRW universe, which has the form
\begin{equation}\label{FL.AP}
dE_{h}=TdS+ W dV,
\end{equation}
where $W=(\rho-P)/2$ is the work density with $\rho$ and $P$ are the
energy density and pressure of the perfect fluid, respectively.
Actually, two kinds of interpretations can be used to understand the
identity~\eqref{FL.AP}. Dynamically, it is the energy balance under
infinitesimal virtual displacements of the horizon normal to itself,
from this perspective, it must be linked with conservation of energy
and thus to the tunneling process. In standard thermodynamics, it is
a connection between two quasi-static equilibrium states of a
system, which differing infinitesimally in the extensive variables
volume, entropy and energy by $dV$, $dS$ and $dE_h$, respectively,
while having same values the intensive variables temperature $T$,
pressure $P$ and density $\rho$. Both of the states are spherically
symmetric solutions of Einstein equations with the radius of horizon
differing by $dr_{h}$ while having the same source $T_{\mu \nu}$.

Correspondingly, the whole setup can be considered from two
different sides. Firstly, as a result of tunneling, some matter
either tunnels out or in across the horizon, therefore energy of the
whole spacetime changes, thus the energy attributed to the shell can
be given out. (Here, one should note that since the two different
quasi-static equilibrium states have the same source $T_{\mu\nu}$,
when the particles radiating out of the horizon, they actrully have
left out of the space we dicussed.) Secondly, considering the
$s$-wave WKB approximation, the imaginary part of the action is
directly related with the Hamiltonian of tunneling particles. Thus,
the first law of thermodynamics is crucial to connect the above two
sides, energy changes of the whole spacetime and the Hamiltonian of
tunneling particles.

The $4$-dimensional FRW metric takes the form
\begin{equation}\label{FRW.metric}
    ds^2=-dt^2+a^2(t)\Big{(}\frac{dr^2}{1-k r^2}+r^2 d\Omega^2_2\Big{)},
\end{equation}
where $t$ is the cosmic time, $r$ is the comoving coordinate, $a(t)$
is the scale factor, $d \Omega^2_2$ represents the line element of a
$2$-dimensional unit sphere, and $k = -1, 0, 1$ is the spatial
curvature constant. The metric \eqref{FRW.metric} can be rewritten
as $ds^2=h_{a b} dx^adx^b +\tilde{r}^2d\Omega^2_2$, with $x^a =(t,
r)$, $h_{a b}={\rm{diag}}(-1, a^2/(1-k r^2))$ and $\tilde{r}=a(t)r$.
By definition $h^{a b}\partial_{a} \tilde{r}
\partial_{b} \tilde{r}=0$, the radius of the locally defined apparent horizon
can be easily given out,
\begin{equation}\label{AH}
    \tilde{r}=\tilde{r}_{A}\equiv \frac{1}{\sqrt{H^2+k/a^2}},
\end{equation}
where $H\equiv \dot{a}/a$ is the Hubble parameter.

It is convenient to use the coordinates $(t, \tilde r)$ to discuss
the tunneling of particles. The metric \eqref{FRW.metric} can be
rewritten as
\begin{equation}\label{T.metric}
d s^{2} = - \frac{1 - \tilde{r}^{2} / \tilde{r}_{A}^{2}}{1 - k
          \tilde{r}^{2} / a^{2}} d t^{2} - \frac{2 H \tilde{r}}{1 - k
          \tilde{r}^{2} / a^{2}} d t d \tilde{r} + \frac{1}{1 - k
          \tilde{r}^{2} / a^{2}} d \tilde{r}^{2} + \tilde{r}^{2} d
          \Omega^{2}_{2}.
\end{equation}
The radial null geodesic ($ds^2=d\Omega^2_2=0$) takes
\begin{equation}\label{null geodesic02}
    \frac{d\tilde{r}}{dt}\equiv \dot{\tilde{r}}
    =H \tilde{r} \pm \sqrt{H^2 {\tilde{r}}^2
    +\Big{(}1-\frac{{\tilde{r}}^2}{{\tilde{r}}_A^2}\Big{)}},
\end{equation}
where $+/-$ corresponding to outgoing/ingoing null geodesics. Since
the observer is inside the apparent horizon, we consider an incoming
mode in the following discussion.

For the definition of surface gravity in such a dynamical systems,
we prefer to Hayward's work~\cite{Hayward}, which is defined as
$K^{b}\nabla_{[a}K_{b]}=\kappa K_{a}$, where $K^a=- \epsilon^{a b}
\nabla_{b} \tilde{r}$ is the Kodama vector corresponding to metric
\eqref{T.metric}, and $\epsilon^{a b}$ denotes the volume form
associate with the $(t, \tilde{r})$ part. The dynamical surface
gravity can be equivalently expressed as
\begin{equation}\label{DSG}
    \kappa=\frac{1}{2}\nabla^{a}\nabla_{b}\tilde{r}.
\end{equation}
Using the metric \eqref{T.metric}, one can find the surface gravity
takes$^{[13]}$
\begin{equation}\label{surface gravity02}
    \kappa=-\frac{1}{\tilde{r}_A}\Big{(}1-\frac{\dot{\tilde{r}}_A}{2 H
    \tilde{r}_A}\Big{)}.
\end{equation}
From Eq. \eqref{null geodesic02} and Eq. \eqref{surface gravity02},
near the apparent horizon, we have
\begin{equation}\label{NAH}
   \dot{\tilde{r}}=-\kappa (\tilde{r}-\tilde{r}_{A}) \frac{1}{H \tilde{r}_{A}} \Big{(}1-\frac{\dot{\tilde{r}}_A}{2 H
    \tilde{r}_A}\Big{)}^{-1}.
\end{equation}

The imaginary part of the action for an $s$-wave ingoing positive
energy particle which crosses the horizon inwards from $\tilde{r}_i$
to $\tilde{r}_f$ can be expressed as
\begin{equation}\label{ingoing action}
    {\rm{Im}} \mathcal{S}
    ={\rm{Im}} \overset{\tilde{r}_{f}}{\underset{\tilde{r}_{i}}{\int}} p_{\tilde{r}}d\tilde{r}
    ={\rm{Im}} \overset{\tilde{r}_{f}}{\underset{\tilde{r}_{i}}{\int}}
    \overset{p_{\tilde{r}}}{\underset{0}{\int}}dp^{\prime}_{\tilde{r}}d\tilde{r}
    ={\rm{Im}}\overset{\mathcal{\tilde{H}}_{f}}{\underset{\mathcal{\tilde{H}}_{i}}{\int}}
    \overset{\tilde{r}_{f}}{\underset{\tilde{r}_{i}}{\int}}\frac{d\tilde{r}}{\dot{\tilde{r}}}d
    \mathcal{\tilde{H}}
    =\overset{\mathcal{\tilde{H}}_{f}}{\underset{\mathcal{\tilde{H}}_{i}}{\int}}\frac{d
    \mathcal{\tilde{H}}}{2T} H \tilde{r}_{A}\Big{(}1-\frac{\dot{\tilde{r}}_A}{2 H
    \tilde{r}_A}\Big{)},
\end{equation}
where we have used the Hamilton's equation
$\dot{\tilde{r}}=d\tilde{\mathcal{H}}/dp_{\tilde{r}}|_{\tilde{r}}$,
the relation between Hawking temperature and surface gravity
$T=|\kappa|/2\pi$, and the contour integral at the pole
$\tilde{r}=\tilde{r}_{A}$.

Evaluating the integral \eqref{ingoing action}, the form of the
Hamiltonian $d\tilde{\mathcal{H}}$ is necessary to be given out.
Now, appealing to energy conversation, we turn to the system to
guess the form of $d\tilde{\mathcal{H}}$. However, explicit
time-dependence in a dynamical system, the Hamiltonian is no-longer
equal to the total energy of the system. Luckily, according to
Hayward's work, we still can determine out the relation between the
Hamiltonian and the total energy of the system.

Two conserved currents can be introduced in our spherical dynamical
system. The first is the Kodama vector $K^{a}$, and the
corresponding conserved charge is the area volume
$V={\underset{\sigma}{\int}}K^{a}d\sigma_{a}=4 \pi \tilde{r}^3/3$,
where $d\sigma_{a}$ is the volume form times a future directed unit
normal vector of the space-like hypersurface $\sigma_{a}$. Another
is defined as the energy-momentum density $j^{a}\equiv
T^{a}_{b}K^{b}$ along the Kodama vector, and the conserved charge is
$E=-{\underset{\sigma}{\int}}j^{a}d\sigma_{a}$ which is equal to the
Misner-Sharp energy.

The total energy inside the apparent horizon can be given as
$E_{h}=\tilde{r}_{A}/2$, which is the Misner-Sharp energy with the
radius $\tilde{r}=\tilde{r}_{A}$ of a spherical
system~\cite{Caietc}. The energy outside the region can be expressed
as $E_{m}=-{\underset{\sigma}{\int}}T^{a}_{b}K^{b}d\sigma_{a}$,
where the integration extends from the apparent horizon to infinity.
Thus the total energy of the spacetime can be expressed as
\begin{equation}\label{total energy}
    E_{T}=\frac{\tilde{r}_{A}}{2}
    -{\underset{\sigma}{\int}}T^{a}_{b}K^{b}d\sigma_{a}.
\end{equation}

As a result of tunneling, the parameters (mass, charge, etc.) which
fix the radius of the horizon change, and this further leads to a
change in the radius of the horizon. We can convince that the only
physical change occurring due to the process of tunneling is the
radius of the horizon. The two states have radius $\tilde{r}_{A}$
and $\tilde{r}_{A}+\delta \tilde{r}_{A}$, respectively, but have a
common source $T_{\mu \nu}$ of perfect fluid with nonzero pressure
$P$ and energy density $\rho$ near the apparent horizon.

However, in our case, we would like to emphasize that as a conserved
charge the Misner-Sharp energy is considered from the viewpoint of a
so-called Kodama observer, whose worldlines are the integral curves
of the Kodama vector. For the metric \eqref{T.metric}, the Kodama
vector is $K^{\mu}=(\sqrt{1-\frac{k \tilde{r}^2}{a^2}}, 0, 0, 0)$.
The energy-momentum tensor of the perfect fluid corresponding to the
comoving coordinate \eqref{FRW.metric} has the form, $T_{\mu
\nu}=(\rho+p)U_{\mu}U_{\nu}+p g_{\mu \nu}$. It is easily to find
that the $(0,0)$ component of energy momentum tensor corresponding
to the metric \eqref{T.metric} takes $T^{0}_{0}=-\rho$.

Now, we can give the energy changes between the final and initial
states of the tunneling process, which contributes to the shell from
the viewpoint of a Kodama observer. According to energy
conservation，we have
\begin{align}\label{energy changes}
    d\mathcal{H}=&E^{f}_{T}(\tilde{r}_{A}+d\tilde{r}_{A})-E^{i}_{T}(\tilde{r}_{A})\nonumber\\
    =&\frac{\delta\tilde{r}_{A}}{2}-\left({{\overset{\infty}{\underset{\tilde{r}_{A}+\delta\tilde{r}_{A}}{\int}}}
    -{\overset{\infty}{\underset{\tilde{r}_{A}}{\int}}}}
    \right) T^{0}_{0} K^{0} d \sigma_{0}=\frac{\delta\tilde{r}_{A}}{2}-\left({{\overset{\infty}{\underset{\tilde{r}_{A}+\delta\tilde{r}_{A}}{\int}}}
    -{\overset{\infty}{\underset{\tilde{r}_{A}}{\int}}}}
    \right) T^{0}_{0} K^{a} d \sigma_{a}\nonumber\\
    =&dE_{h}-\rho d V.
\end{align}
Since the energy difference $d\mathcal{H}$ is measured by a Kodama
observer inside the apparent horizon, near the apparent horizon we
have
\begin{equation}\label{our observer}
    d\tilde{\mathcal{H}}=\frac{d\mathcal{H}}{\sqrt{1-\frac{k
    \tilde{r}^2}{a^2}}\mid_{\tilde{r}=\tilde{r}_{A}}}=\frac{d\mathcal{H}}{H \tilde{r}_{A}}
\end{equation}
Substituting Eq. \eqref{energy changes} and Eq. \eqref{our observer}
into Eq. \eqref{ingoing action}, yields
\begin{equation}\label{action relation}
    {\rm{Im}} \mathcal{S}=\overset{\mathcal{H}_{f}}{\underset{\mathcal{H}_{i}}{\int}}\frac{d
    \mathcal{H}}{2T} \Big{(}1-\frac{\dot{\tilde{r}}_A}{2 H
    \tilde{r}_A}\Big{)}
    =\int\frac{dE_{h}-\rho d V}{2T} \Big{(}1-\frac{\dot{\tilde{r}}_A}{2 H
    \tilde{r}_A}\Big{)}.
\end{equation}
The above expression can be further simplified. The Friedmann
equation of spacetime and the continuity equation of the perfect
fluid $T_{\mu\nu}$ have the form
\begin{equation}\label{Frie.Con.}
    H^2+\frac{k}{a^2}=\frac{8\pi}{3}\rho,~ ~ ~ \dot{\rho}+3H(\rho+P)=0.
\end{equation}
Using Eq. \eqref{AH} and the Friedmann equation, one can easily
check that the total energy inside the apparent horizon satisfies
$E_{h}=\tilde{r}_{A}/2= \rho V$ with $V=4\pi \tilde{r}_{A}^3/3$.
Thus, we have
\begin{equation}\label{E.further expr.}
    dE_{h}=d(\rho V)= V \dot{\rho} dt+\rho
    dV=-3H (\rho+P)V dt+\rho dV.
\end{equation}
Combining Eq. \eqref{action relation} and Eq. \eqref{E.further
expr.}, one can obtain
\begin{equation}\label{action relation Simp.}
    {\rm{Im}} \mathcal{S}=\int\frac{dE_{h}-W d V}{2T},
\end{equation}
where $W=(\rho-P)/2$ is the work density.

Using the unified first law of thermodynamics Eq. \eqref{FL.AP},
from Eq. \eqref{action relation Simp.} we finally have
\begin{equation}\label{finally equation}
    {\rm{Im}} \mathcal{S}=\int\frac{dS}{2}.
\end{equation}
Now, one can immediately obtain the semi classical tunneling rate in
the FRW universe,
\begin{equation}\label{tunneling rate}
   \Gamma \sim e^{-2 \texttt{Im} \mathcal{S}}=e^{-\int^{S_{f}}_{S_{i}}dS
}=e^{-\triangle S},
\end{equation}
where $\triangle S= S_{f}-S_{i}$. This is the well-known result
obtained in Ref.~\cite{Parikh} for a general, stationary,
asymptotically flat, spherically symmetric black hole background.
And as a consequence of the first law of thermodynamics, this result
appears in Refs.~\cite{Sarkar, Pilling}. Here, we recover it in a
dynamical spacetime background, the FRW universe. The minus signs
which appear in the last two exponential index in Eq.
\eqref{tunneling rate} causes by the entropy differences of initial
and final states of the system, here obviously we have $S_{f} >
S_{i}$.

In summary, we analyze how tunneling is intimately connected with
the unified first law of thermodynamics through the principle of
energy conversation in a dynamical system, the FRW universe. Our
discussion shows that the tunneling rate arises as a natural
consequence of the unified first law of thermodynamics $dE_{h}=TdS+
W dV$ at the apparent horizon.

On the other hand, our proceeding clearly expresses the physical
meaning of the locally defined Hawking temperature
$T=|\kappa|/2\pi$, which is associated with the apparent horizon of
the FRW universe. For such a dynamical system, the Hawking
temperature is measured by an Kodama observer inside the horizon.

The author (Jiang K.-X.) would like to thank Dr. Zhu T. for pointing
out some defects and helpful discussion.


\end{document}